# Edge optical scattering of two-dimensional materials

*Huaiyi Ding[a, b], Yiyun Dong[a], Sijia Li[a], Nan Pan[a, b, c]\*, Yi Luo[a, b], Xiaoping Wang[a, b, c] \**

[a] Hefei National Laboratory for Physical Sciences at the Microscale, University of Science and Technology of China, Hefei, Anhui 230026, P. R. China.

[b] Synergetic Innovation Center of Quantum Information & Quantum Physics, University of Science and Technology of China, Hefei, Anhui 230026, P. R. China

[c] Key Laboratory of Strongly-Coupled Quantum Matter Physics, Chinese Academy of Sciences, School of Physical Sciences, University of Science and Technology of China, Hefei, Anhui 230026, P. R. China

*Correspondence to Nan Pan (npan@ustc.edu.cn)

Rayleigh scattering has shown powerful abilities to study electron resonances of nanomaterials regardless of the specific shapes. In analogy to Rayleigh scattering, here we demonstrate that edge optical scattering from two-dimensional (2D) materials also has the similar advantage. Our result shows that, in visible spectral range, as long as the lateral size of a 2D sample is larger than 2 μm, its edge scattering is essentially a knife-edge diffraction, and the intensity distribution of the high-angle scattering in $k$ space is nearly independent of the lateral size and the shape of the 2D samples. The high-angle edge scattering spectra are purely determined by the intrinsic dielectric properties of the 2D materials. As an example, we experimentally verify this feature in single-layer MoS$_2$, in which A and B excitons are clearly detected in the edge scattering spectra, and the scattering images in $k$ space and real space are consistent with our theoretical model. This study shows that the edge scattering is a highly practical and efficient method for optical studies of various 2D materials as well as thin films with clear edges.

## I Introduction

Rayleigh scattering, an elastic scattering of light when particle size is much smaller than the wavelength of light, reflects the contrast of dielectric constants between the particle and its background regardless of its shape. It is much brighter than inelastic scattering signals, such as Raman scattering and photoluminescence, and possesses a much higher contrast than that of reflection and transmission images. Due to these unique advantages, Rayleigh scattering has become a powerful method for nanoparticle imaging [1] and disordered potential studies in two-dimensional (2D) electron gas [2,3]. Traditionally, it is often considered that Rayleigh scattering can only be applied to zero-dimensional (0D) samples. But recently, it was successfully applied to one-dimensional (1D) materials in which the scattered light has the similar behaviors to 0D cases. One of the most successful applications is the optical studies of individual single-walled carbon nanotubes, such as chiral index identification [4-6], optical imaging [7], excitonic optical wires study [8], and interaction between nanotubes [9]. In principle, further generalizing the Rayleigh scattering method to 2D structures is infeasible, because 2D coherent dipoles excited by the incident light can only generate reflected and transmitted light. For this reason, to the best of our knowledge, the dielectric properties of 2D materials can only be measured by reflection or transmission spectra, such as ellipsometry [10-13] and differential reflection method [14]. Edges of 2D materials are considered to be another kind of structure other than 1D and 2D ones, which often possess significant and unique physical properties different from bulks [15-19]. Unfortunately, similar to 0D and 1D structures, the optical reflection and transmission measurements of the edges suffer from the low response compared with the background. However, the edge also has scattering signal because the broken translational symmetry provides effective momentum compensation necessary for the light scattering. In this paper, we show that the high-angle edge scattering signal is exclusively determined by the intrinsic dielectric properties of the sample's edge. Furthermore, in visible spectral range, it is approximately independent of the lateral size and the shape of the 2D sample as long as its lateral size is larger than 2 μm (~4λ, where λ is the wavelength of the scattered light). This requirement makes edge scattering nicely suitable for almost all 2D samples that could be obtained in laboratories. We also demonstrate an experimental setup for high-angle edge optical scattering measurement in real space and $k$ space. Under this setup, A and B excitons' resonance peaks are unambiguously resolved from the edge of monolayer $MoS_2$ flakes.

Our results show that the edge scattering could be a powerful and convenient method for the optical studies of various 2D materials' edges as well as thin films with clear edges. It is also promising for a variety of characterizations in 2D systems such as grain/domain boundaries, phase separation/transition, interlayer coupling, topological edge states, heterojunctions, and so on, as long as the discontinuity of dielectric properties exists.

## II Theoretical result

The theoretical model considers a ribbon sample with width of 2a (Fig. 1(a)) and infinite length, which is placed on a flat substrate. The thickness of the ribbon is less than one-tenth of the incident light wavelength and thus has no effect to the scattering in this model. A parallel beam of light with normal incidence to the substrate covers the sample entirely. The difference between the dielectric constants of the sample and the substrate leads to the corresponding discontinuity of the dipole field. We consider the contribution of the dipole field of the sample to the outgoing light in two parts. One part of the contribution is equal to the effect of the substrate, which includes reflected and transmitted light. The rest part of the contribution is the scattered light from the sample, which is related to the difference between the dipole field of the sample and the substrate, as indicated in Fig. 1(a). Far-field radiation generated by this contribution of the dipole field of the sample can be analogized to single-slit diffraction with slit width of $2a$. When the width of the ribbon sample is infinite, the edge scattering observed at an edge of the sample is equivalent to the knife-edge diffraction. In order to simplify the calculation, we first consider the dipole moments of the sample and the substrate as a scalar. The distribution of far-field radiation in $\mathbf{k}$ space can be derived from the Fourier transform of the dipole field. The strength distribution of the difference between the dipole field of the sample and the substrate can be written as a scalar field function $D(x)$,

$$D(x) = \begin{cases} 0 & |x| > a \\ A & |x| < a \end{cases}, \quad (1)$$

where $A$ is the complex amplitude of the field. The dipole field is translational invariant in y direction, thus is $y$ independent. If the dielectric constant of the sample is much larger than that of the substrate, $A = (\varepsilon_{sample} - \varepsilon_{substrate})E \approx \chi E$, where $\varepsilon_{sample}$ and $\varepsilon_{substrate}$ are the dielectric constants of the sample and the substrate respectively, $E$ is the complex amplitude of the incident light field, and $\chi$ is the electric susceptibility of the sample. When the light is normal incident to the sample, the momentum of the dipole field in x-y plane is zero and $A$ is independent of the spatial coordinates. Fourier transform of $D(x)$ is

$$F(k) = \frac{1}{2\pi}\int_{-\infty}^{+\infty} D(x)e^{-ikx}dx = \frac{A}{2\pi}\int_{-a}^{+a} e^{-ikx}dx = \frac{A}{\pi}\frac{\sin(ka)}{k}, \qquad (2)$$

which is plotted in Fig. 1(b). Since the far-field radiation intensity is proportional to the square of the dipole oscillation frequency and its intensity, the distribution of the electric field strength in $\boldsymbol{k}$ space, $E(k)$, is proportional to $F(k)$

$$E(k) = \gamma\omega^2 F(k) = \gamma\omega^2 \frac{A}{\pi}\frac{\sin(ka)}{k}, \qquad (3)$$

where $\omega$ is the angular frequency of the electric field, $\gamma$ is a constant related to dipole radiation. It can be considered as a sine oscillation function with 1/k amplitude modulation. $Sin(ka)/k$ function is plotted in Fig.1(b) while $a$=5 and 10$\mu$m. When k is equal to 0, there is a main peak with the amplitude of $Aa/\pi$, which contributes to back and forward scattering. The other peaks are slightly affected by the sample width parameter $a$, including the part of our attention that generates high-angle scattering. Intuitively, high-angle scattering intensity does not depend on the lateral size. Especially when $a$ is large enough, $F(k)$ becomes a high-frequency oscillation in $\boldsymbol{k}$ space and therefore can be replaced by a mean value. The intensity of high-angle scattering is given by the integration of the range ($k_1$, $k_2$) where $k_1$, $k_2$>0,

$$I_{(k_1,k_2)} = \int_{k_1}^{k_2} |E(k)|^2 dk = \left(\gamma\omega^2 \frac{A}{\pi}\right)^2 \int_{k_1}^{k_2} \frac{\sin^2(ka)}{k^2} dk$$

$$\approx \left(\gamma\omega^2 \frac{A}{\pi}\right)^2 \int_{k_1}^{k_2} \frac{1}{2k^2} dk = \frac{1}{2}\left(\frac{1}{\pi}\right)^2 (\gamma\omega^2 \chi E)^2 \left(\frac{1}{k_1} - \frac{1}{k_2}\right), \qquad (4)$$

where $\sin^2(ka)$ is approximately 1/2. It is clearly that $I_{(k_1,k_2)}$ after approximation is independent of the lateral size 2$a$. Taking the integration interval as (10$\mu$m$^{-1}$, 20$\mu$m$^{-1}$), we compare the numerical integrated result without approximation with the integration after approximation in Fig. 1(c). As $a$ increases, the intensity without approximation gradually converges to the intensity after approximation. When $a>1\mu m$, the difference between two results is less than 5%. This means that the intensity of high-angle scattering gradually become independent of the lateral size 2 $a$ as $a$ increases.

Next we demonstrate that the far-field high-angle scattering is mainly induced by the edge. It can be obtained by its high-angle real space image, that is the inverse Fourier transform of $F(k)$ in range $(k_1, k_2)$,

$$R_{(k_1,k_2)}(x) = \gamma\omega^2 \int_{k_1}^{k_2} F(k)e^{ikx}dk = \gamma\omega^2 \chi E \int_{k_1}^{k_2} \frac{\sin(ka)}{\pi k} e^{ikx}dk. \qquad (5)$$

For the finite integration interval $(k_1, k_2)$, there is no analytical solution to Eq. (5). Integrating over the range $(10\mu m^{-1}, 20\mu m^{-1})$, we give the numerical integration of $\left|\int_{k_1}^{k_2} \frac{\sin(ka)}{k} e^{ikx} dk\right|$ for $a=1, 2.5, 5, 10$ and $15\mu m$ in Fig. 1(d), while the integrated results of other integrating range are shown in Supplemental Materials [20]. As indicated in Fig. 1(d), most of the high-angle scattered light concentrates on the edge in real space image. In accord with the result of Eq. (4) and Fig. 1(c), the intensity of the high-angle scattering at the edge is nearly independent of the lateral size. It's also demonstrated in Fig. 1(e) that the phase distribution of $R_{(k_1,k_2)}(x-a)$ around the edge is independent of the lateral size $2a$. Therefore we can conclude that the dipole field in real space $R_{(k_1,k_2)}(x)$ around the straight edge is weakly affected by the lateral size and shape of the sample. This means that in practical applications, similar to Rayleigh scattering method, it is generally not necessary to consider shape and size factors when extracting the dielectric information from the high-angle scattering spectra.

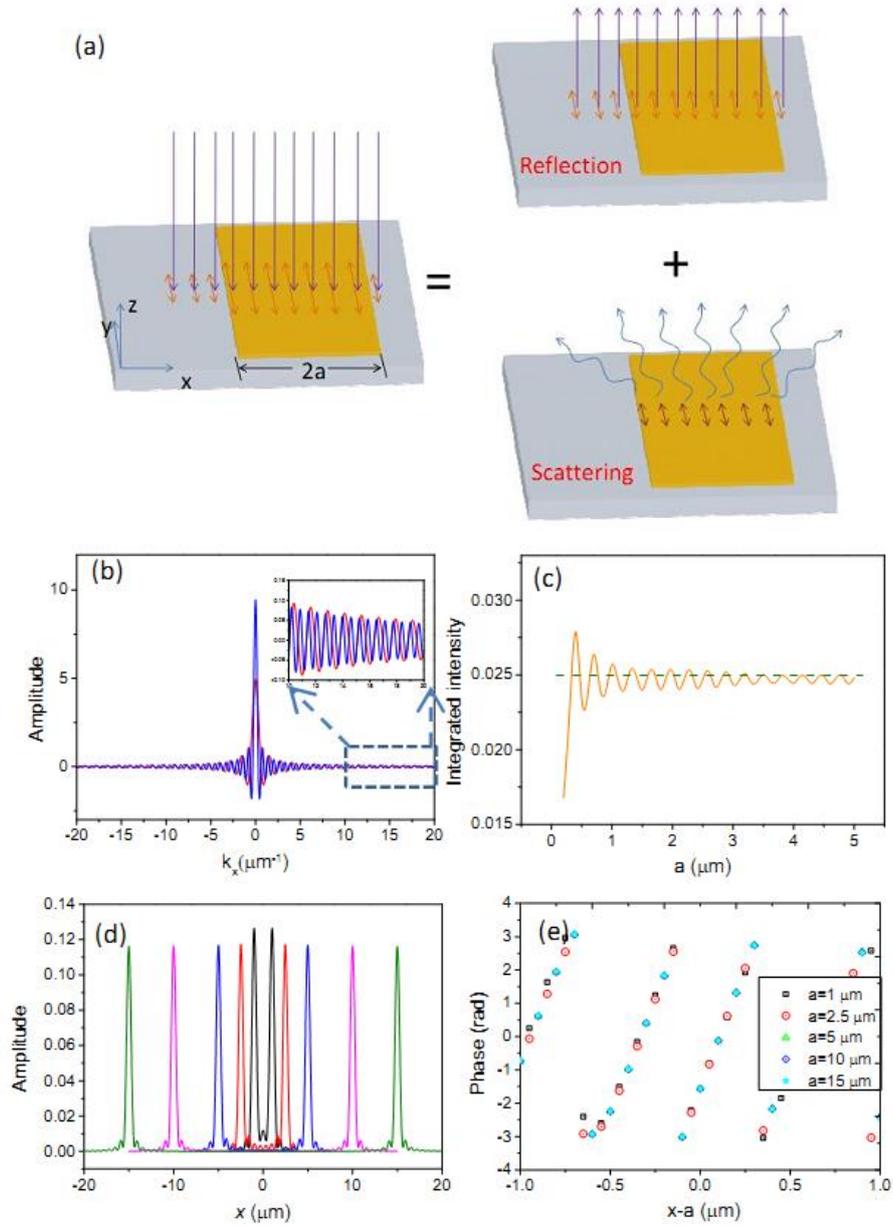

FIG. 1. (a) Decomposition of the dipole field. The dipole field (orange arrow) in left graph can be divided to two parts: the reflection part as shown in the upper right graph, and the scattering part in the lower right graph. (b) Scattering distribution in $\boldsymbol{k}$ space when lateral size $2a$=10 (red line) and 20 μm (blue line), respectively. Insert is the enlarged curve of the dashed box in (b). (c) The numerical (orange line) and approximate (green dot line) integrated results as a function of the lateral size $2a$. For simplicity, these curves are normalized to the integrated intensity of the approximate result. The integrating range is (10 $\mu m^{-1}$, 20 $\mu m^{-1}$). (d) The intensity distribution of high-angle scattering in real space for different lateral sizes: black, red, blue, pink and green denotes the case for $a$=1, 2.5, 5, 10, and 15 μm, respectively. The integrating range is also (10

μm⁻¹, 20 μm⁻¹). (e) Phase distribution of $R_{(k_1,k_2)}(x-a)$ with different lateral sizes.

In the theoretical model, we considered the scenario of high-angle scattering under normal incidence. However, for convenience, we use grazing incident beam in experiment so that the collection of the high-angle scattering can be almost normal, as illustrated in Fig. 2(a). The incident beam is in the *x-z* plane, with incidence angle $\theta$ larger than *arcsin(N.A.)*, where *N.A.* is the numerical aperture of objective lens. The reflected light and forward scattering is out of the range of objective lens *N.A.*, and thus only high-angle scattering can be collected. If the edge is perpendicular to the incident plane as in Fig. 2(b), reflected and scattered light is all in the *x-z* plane. In the image of **k** space described in Fig. 2(c), the wave vector of the reflected light in the $(k_x, k_y)$ plane is $(k_0 \sin\theta, 0)$ and the scattering is a line along $k_x$ axis. Due to the non-zero incident angle, the incident beam has a different phase at different locations of the sample and the substrate. A phase term should be added to the previously defined scalar field function $D(x)$,

$$D(x) = \begin{cases} 0 & |x| > a \\ Ae^{-ik_0 x \sin\theta} & |x| < a \end{cases}, \quad (6)$$

After a variable substitution $k_{new} = k + k_0 \sin\theta$, the Fourier transform of $D(x)$ can be the same as Eq. (2). Since the collection range of objective lens is $(-k_0 N.A., k_0 N.A.)$, the integrating interval of the high-angle scattering intensity as Eq. (4) becomes $(k_0 \sin\theta - k_0 N.A., k_0 \sin\theta + k_0 N.A.)$ after the variable substitution. In practice the incident plane is fixed in the *x-z* plane, however, the flakes are randomly oriented on the substrate so there is always a non-zero angle between each edge of those flakes and the *y* axis as shown in Fig. 2(d). Therefore, the scattering obtained above should be further generalized. Considering an edge with an infinite length which has a non-zero angle $\alpha$ to the *y* axis, the edge scattering in **k** space must be perpendicular to the edge in real space, which is similar to one-dimensional Rayleigh scattering in Ref. [8]. The wave vector of the reflected light in the $(k_x, k_y)$ plane is $(k_0 \sin\theta, 0)$, regardless of the orientation of the edge. Fig. 2(e) shows the illustration of the scattering distribution in **k** space. Because the collection area of the objective lens is the area of a circle with center on the origin and radius $k_0 N.A.$, the integrating

interval of the high-angle scattering intensity is

$$(k_0 \sin\theta \cos\alpha - k_0\sqrt{N.A.^2 - (\sin\theta \sin\theta)^2}, k_0 \sin\theta \cos\alpha + k_0\sqrt{N.A.^2 - (\sin\theta \sin\theta)^2}), \quad (7)$$

in this configuration. And the distance between the origin and scattering line is $d = k_0 \sin\theta \sin\alpha$. Fourier transformation analysis of the 2D dipole field,

$$F(k_x, k_y) = \left(\frac{1}{2\pi}\right)^2 \iint D(x,y) e^{-i(k_x x + k_y y)} dx dy, \quad (8)$$

can also give this result, which is further derived in Supplemental Material [20]. The integrating interval of the high-angle scattering intensity in Eq. (7) also indicates that the angle $|\alpha|$ between the edge and the $y$ axis must be less than $\alpha_c = arcsin(N.A./sin\theta)$, otherwise the edge scattering will be out of the objective lens collecting range, as shown in Fig. 2(e).

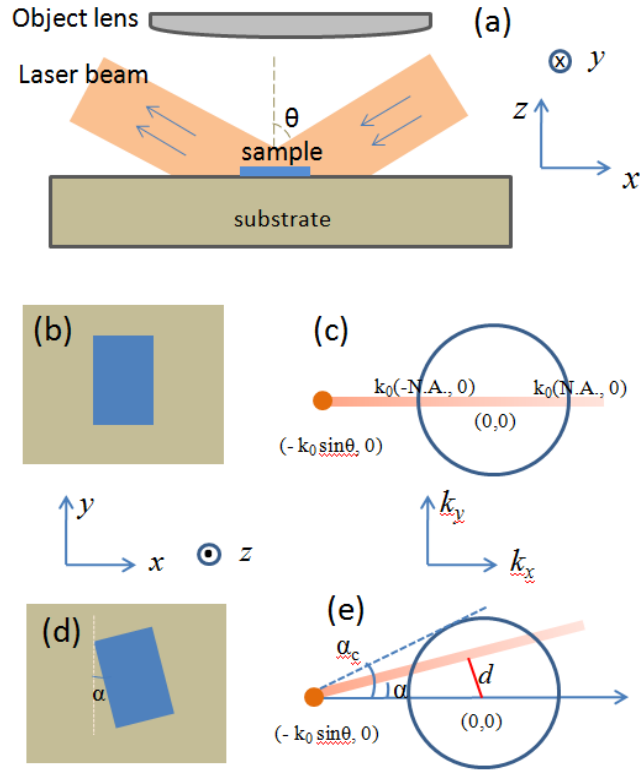

FIG. 2. Illustration of edge scattering microscopy of 2D samples with different orientations. (a) Experimental setup and the definition of coordinate system. (b) Sample's edge which parallels to $y$ axis. (c) Sketch of the scattering distribution in $\boldsymbol{k}$ space corresponding to the sample in (b). Blue circle is the collection area of the objective lens. (d) Sample's edge which has an included angle $\alpha$ to $y$ axis. (e) Sketch of the scattering distribution in $\boldsymbol{k}$ space corresponding to the sample

in (d).

So far we have not considered the dipole moment of the sample as a vector. The Fourier transform method can accurately describe the angular distribution of the radiation from a non-polarized dipole field, but for a polarized or partially polarized dipole field, their intensity distribution in $k$ space is related to the dipole polarization direction. Thus, a correction factor should be added to the intensity distribution in $k$ space as written below,

$$I_c(k_x, k_y) = I_V(k_x, k_y)|E(k_x, k_y)|^2, \qquad (9)$$

$$I_V(k_x, k_y) = \left|\frac{\vec{p}}{|\vec{p}|} \times \frac{\vec{k}}{|\vec{k}|}\right|^2 = |(k_x^2 + k_y^2)p_z^2 + (k_y^2 + k_z^2)p_x^2 + (k_z^2 + k_x^2)p_y^2 - 2(p_x p_y k_x k_y + p_y p_z k_y k_z + p_z p_x k_z k_x)|/\left(|\vec{p}|^2|\vec{k}|^2\right), (10)$$

where $I_C(k_x,k_y)$ is the corrected intensity distribution in $k$ space of the high-angle scattering of the dipole field, and $I_V(k_x,k_y)$ is the correction factor reflecting the influence of dipole polarization. The components of $k$ satisfy $k_x^2 + k_y^2 + k_z^2 = k_0^2$, where $k_0$ is the wave vector of the light in vacuum and $\vec{p}$ is the electric polarization density which is in proportion to the dipole moment. For some particular cases, this equation can be simplified. For example, when the dipole field is non-polarized, the mean value of each cross term, including $p_x p_y$, $p_y p_z$ and $p_z p_x$, is zero, and the square terms $p_x^2$, $p_y^2$ and $p_z^2$ are respectively equal to $|\vec{p}|^2/2$. It results in $I_V(k_x,k_y)=1$, meaning the correction factor can be omitted. Non-polarized fluorescence belongs to this case. However, in our experiment, this correction factor must be considered because the incident light is polarized. Different polarization directions of the incident light cause different values of the sample's electric polarization density. Here we consider s wave and p wave respectively. When the electric susceptibility $\chi$ of the sample is isotropic,

$$\begin{aligned}\vec{p}_s &= (0, \chi E_s, 0) \\ \vec{p}_p &= (\chi E_p \cos\theta, 0, \chi E_p \sin\theta)\end{aligned} \qquad (11)$$

Then,

$$I_{Vs}(k_x, k_y) = \left(1 - \frac{k_y^2}{k_0^2}\right), \qquad (12)$$

$$I_{Vp}(k_x, k_y) = \left(1 - \frac{k_x^2}{k_0^2}\right)\cos^2\theta + \left(1 - \frac{k_z^2}{k_0^2}\right)\sin^2\theta - 2\cos\theta\sin\theta\frac{k_x k_z}{k_0^2}$$

$$= 1 - \left(\frac{k_x}{k_0}\cos\theta + \frac{k_z}{k_0}\sin\theta\right)^2. \qquad (13)$$

If the electric susceptibility $\chi$ of the sample is anisotropic, $\chi$ is a complex tensor

$$\chi = \begin{pmatrix} \chi_{xx} & \chi_{xy} & \chi_{xz} \\ \chi_{yx} & \chi_{yy} & \chi_{yz} \\ \chi_{zx} & \chi_{zy} & \chi_{zz} \end{pmatrix}. \qquad (14)$$

Here we discuss the scenario where the material has a uniaxial crystal structure, which is the case for a wide range of 2D materials are uniaxial crystals, such as MoS$_2$. Their electric susceptibility can be written as the complex tensor below,

$$\chi = \begin{pmatrix} \chi_{in} & 0 & 0 \\ 0 & \chi_{in} & 0 \\ 0 & 0 & \chi_{out} \end{pmatrix}, \qquad (15)$$

where the coordinate system is defined as in Fig. 2(a), and $\chi_{in}$ and $\chi_{out}$ are the in-plane and out-of-plane complex electric susceptibility respectively. Then the dipole can be written as

$$\vec{p} = \chi \vec{E} = (\chi_{in} E_x, \chi_{in} E_y, \chi_{out} E_z). \qquad (16)$$

When the edge scattering is excited by s-polarized light, $\vec{p} = (0, \chi_{in} E_s, 0)$, then

$$I_{Vs}(k_x, k_y) == \left(1 - \frac{k_y^2}{k_0^2}\right), \qquad (17)$$

When the edge scattering is excited by p-polarized light, $\vec{p} = (\chi_{in} E_p \cos\theta, 0, \chi_{out} E_p \sin\theta)$, then

$$I_{Vp}(k_x, k_y) = \left| \frac{(\chi_{in}\cos\theta)^2(k_0^2 - k_x^2) + (\chi_{out}\sin\theta)^2(k_x^2 + k_y^2) + \chi_{in}\chi_{out}\sin\theta\cos\theta}{((\chi_{in}\cos\theta)^2 + (\chi_{out}\sin\theta)^2)k_0^2} \right|$$

$$= \left|1 - \frac{(k_x \chi_{in}\cos\theta + k_z \chi_{out}\sin\theta)^2}{((\chi_{in}\cos\theta)^2 + (\chi_{out}\sin\theta)^2)k_0^2}\right| \qquad (18)$$

Then

$$I_c(k_x, k_y) = I_V(k_x, k_y)|E(k_x, k_y)|^2 =$$

$$\begin{cases} \gamma\omega^2\left(1 - \frac{k_y^2}{k_0^2}\right)|\chi_{in} E_s F'(k_x, k_y)|^2 & s\text{ wave} \\ \gamma\omega^2\left|(\chi_{in}\cos\theta)^2 + (\chi_{out}\sin\theta)^2 - \left(\frac{k_x}{k_0}\chi_{in}\cos\theta + \frac{k_z}{k_0}\chi_{out}\sin\theta\right)^2\right| E_p^2 F'^2(k_x, k_y)| & p\text{ wave} \end{cases}$$

$$, \qquad (19)$$

where $F'(k_x, k_y) = F(k_x, k_y)/|\chi E|$. It's easy to obtain the spectra of $|\chi_{in}|^2$ from the edge scattering excited by s wave. Deriving $\chi_{out}$ is more complicated, because the edge scattering excited by p wave is determined by two complex numbers $\chi_{in}$ and $\chi_{out}$. Therefore, in principle $\chi_{in}$ and $\chi_{out}$ can be derived from the experiment data with the scattering distribution in $k$ space described in Eq. (19).

## III Experimental result

In our experiment, CVD-grown regular triangular flakes of monolayer MoS$_2$ on a SiO$_2$ substrate are served as the samples. *N.A.* of the objective lens is 0.5, corresponding to a ±30º collection angle. The incident angle of light, generated by a SuperK EXR-15 white laser, is set to *θ*=66º (much larger than 30º), thus the reflected light is completely out of collection range of the objective lens. The angle between the edge and the *y* axis (the line perpendicular to the incident plane) should be no more than a critical angle $α_c$=*arcsin*(0.5/*sin*66º)=33º. Since the inner angle of the regular triangle is 60º, which is about twice of the critical angle $α_c$, it is natural to observe only one of the three edges in the scattering images in most of our cases. Fig. 3(a) shows the typical optical microscope image of the samples. The corresponding edge scattering image is shown in Fig. 3(b), besides the greatly enhanced contrast, one can also find that no more than one edge in each flake is "bright". To further study the scattering distribution in $k$ space, an edge of a specific flake is studied by the Fourier transform system with the help of an aperture in real space. The optical path diagram is detailed in Supplemental Material [20]. Fig. 3(c) shows the intensity distribution image in real space of the edge scattering from a single flake after the aperture, and Fig. 3(d) is the corresponding intensity distribution image in $\tilde{k}$ space, where $\tilde{k} = \vec{k}/k_0$. It is clear that the scattering lines in Fig. 3(c) and Fig. 3(d) are perpendicular to each other in real space and $\tilde{k}$ space, with the same rotation angle *α*=17º respectively to the *y* axis and to the *x* axis. As shown in Figs. 3(d), the distance between the origin and the scattering line in $k$ space is $\tilde{d}$=0.266, perfectly matching the theoretical value of $\tilde{d} = \frac{d}{k_0} = \sin(66º)\sin(17º)=0.267$. In addition, the collected intensity of the scattering line in $k$ space gradually decreases from left to right, consistent with the attenuated form of the approximate integrated scattering ($\propto 1/k^2$ as in Eq. (4)). Other samples with different rotation angles also have the same edge scattering properties, the details can be found in Supplemental Material [20].

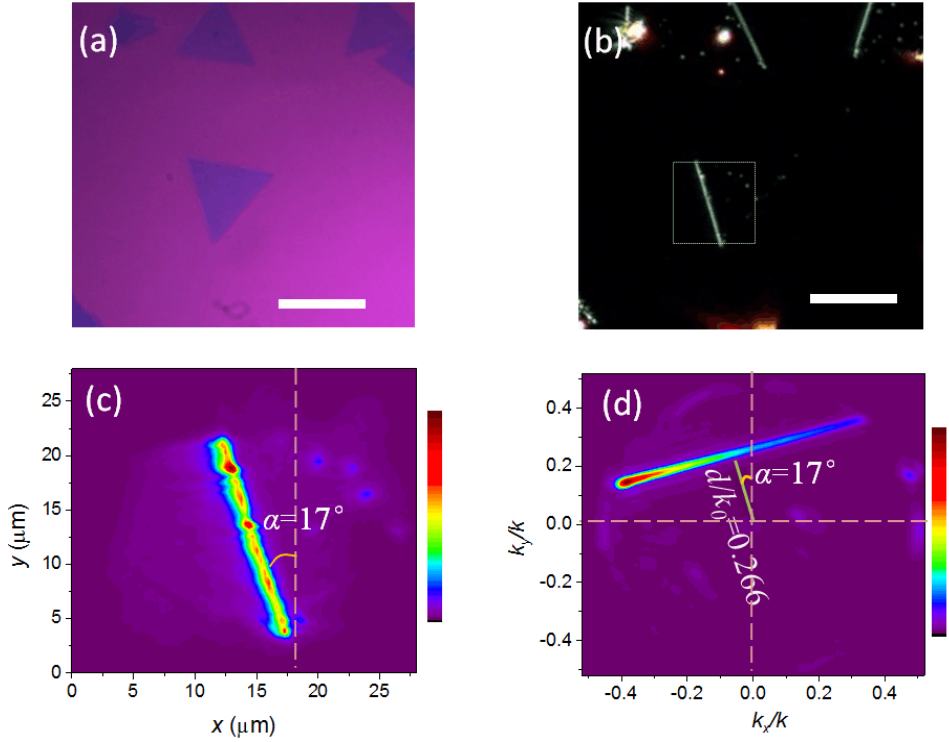

FIG. 3. Experimental results of the edge optical scattering from monolayer MoS$_2$ flakes. (a) Reflection image of several monolayer flakes. (b) The corresponding edge scattering image. Dashed box indicates the aperture allowed accessing area in real space. The scale bar is 20 μm. (c) and (d): The high-angle scattering image of a flake after the aperture (c) in real space and (d) in $\tilde{k}$ space.

The edge scattering spectra can be directly collected by a spectrometer. An example spectrum is shown in Fig. 4(a), with the optical path factors, such as the incident light spectra and the optical loss spectra of the optical elements been corrected. It is clear that A and B exciton peaks are unambiguously resolved in both s wave and p wave scattering spectra. The $|\chi_{in}(\omega)|^2$ is shown in Fig. 4(b), which is derived from the s wave scattering spectrum in Fig. 4(a) with the scattering function in Eq. (19).

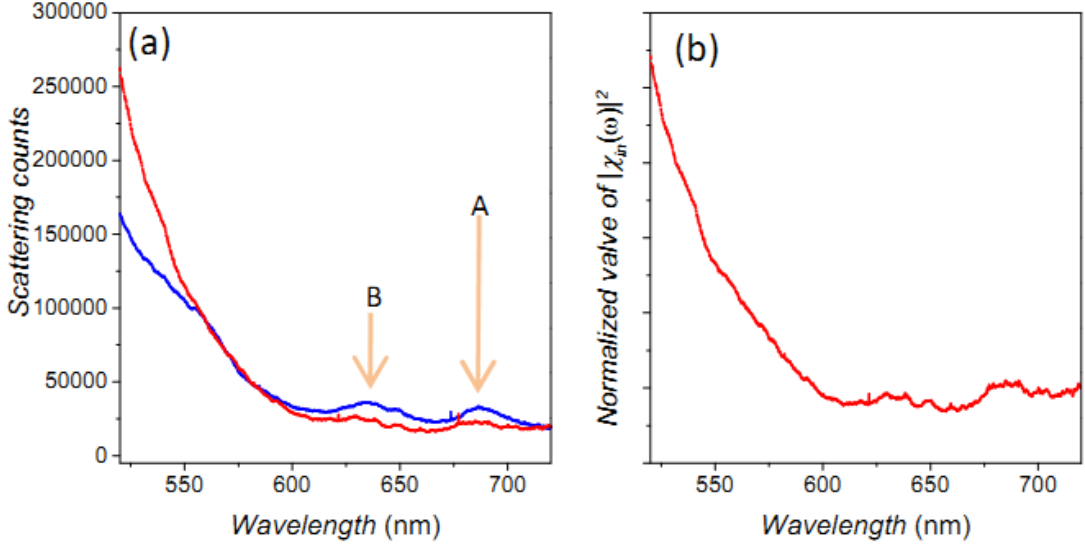

FIG. 4. (a) s wave (red line) and p wave (blue line) edge scattering spectra from a single flake of monolayer MoS$_2$. (b) Normalized value of $|\chi_{in}(\omega)|^2$.

## Ⅳ Conclusion

At last, we discuss the spatial resolution of this high-angle edge scattering image. Since it is a kind of far-field images, the full width at half maximum (FWHM) of an isolated edge's scattering line in real space is limited by *N.A.* of the objective lens, which equals to *λ/N.A.*, regardless of the incident angle. In this case, the edge scattering reflects the dielectric properties near the edge with a scale of *λ/N.A.* However, if the distance between two edges of a sample is less than the optical diffraction limit, those two edges can still be clearly resolved with the larger $k_x$, such as in evanescent wave excitation [21]. Therefore in this case, the edge scattering could also have a super resolution beyond the optical diffraction limit, which could be implemented to resolve ultra-narrow gaps of 2D materials.

In conclusion, we theoretically demonstrate that the high-angle edge optical scattering of 2D materials is independent of its lateral size and shape, and hence the dielectric properties of the materials are reflected. We also show that the scattering lines in real space and ***k*** space are perpendicular to each other, which is similar to Rayleigh scattering of 1D materials. It is experimentally demonstrated by scattering image of the edges of monolayer $MoS_2$ flakes in spatial and angle resolved optical system. Its A and B excitons' peaks are unambiguously resolved in scattering spectra. These results illustrate that high-angle edge scattering is a powerful method for the study of the edges' optical properties of 2D materials.

Acknowledgement

We acknowledge the financial supports from MOST of China(2016YFA0200602), National Natural Science Foundation of China (21421063, 11474260, 11374274, 11504364, 21633007), the Chinese Academy of Sciences (XDB01020200),the Fundamental Research Funds for the Central Universities (WK2030020027, WK2060190027, WK3510000004), and Anhui Natural Science Foundation (1608085QA17).

Competing financial interests: The authors declare no competing financial interests.

# Supplemental Material for "Edge optical scattering of two-dimensional materials"

**Section A: Inverse Fourier transform in different areas of *k* space**

The approximation of Eq. (4) in main text can only be satisfied when integration interval ($k_1$, $k_2$) is far away from the origin, or namely high-angle area. We calculate the edge scattering amplitude of different integration interval to determine which area in k space is suitable. For *N.A.*=0.5 objective lens, the diameter of collecting range in k space is 10μm$^{-1}$ at 628nm. Then all the integration length $|k_2-k_1|$ is set as 10μm$^{-1}$ to be according with the experiment parameters. From Fig. S1 we can see that, in low angle integration area (1μm$^{-1}$, 11μm$^{-1}$), scattering light amplitude highly depends on the sample lateral size. While $k_1$>3μm$^{-1}$, Fig. S1(e-i), amplitude fluctuation can be nearly neglected for the samples with deferent width. For the experiment parameters in main text, oblique incident angle of exciting laser *θ*=66º and numerical aperture of object lens *N.A.*=0.5, the integration area range from (5.2μm$^{-1}$, 17.8μm$^{-1}$) to (3.7μm$^{-1}$, 12.7μm$^{-1}$) at (500nm, 700nm) range wavelength. They are all in the approximate reasonable region.

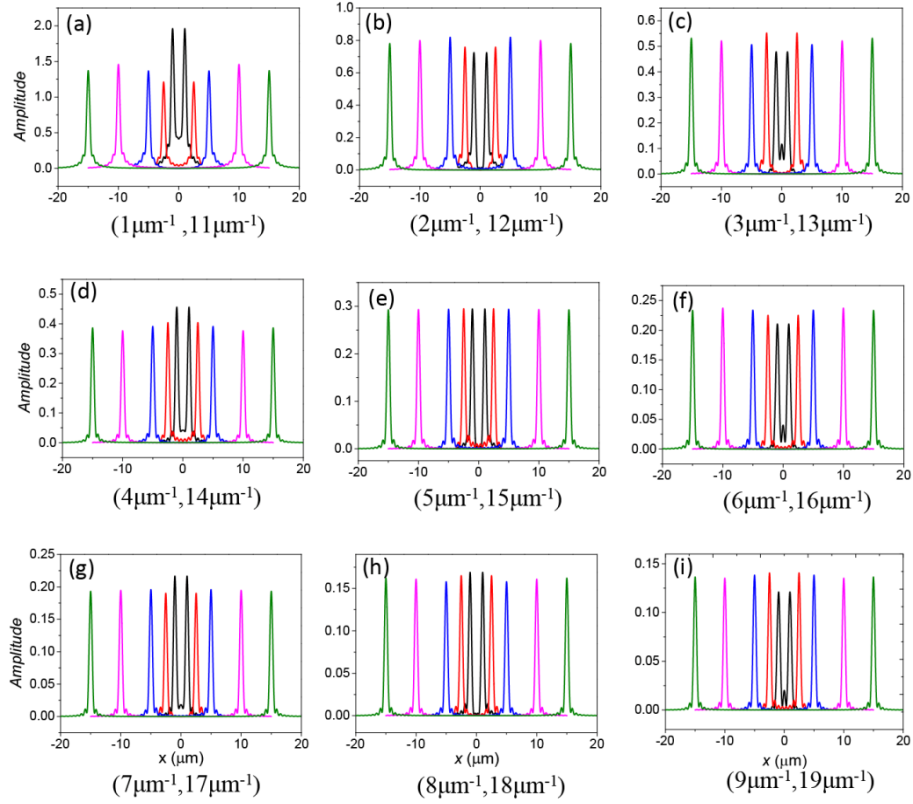

FIG. S1. Amplitude of inverse Fourier transform of $\frac{sin(ka)}{k}$ with different integration interval and sample lateral size, black $a=1\mu m$, red $a=2.5\mu m$, blue $a=5\mu m$, pink $a=10\mu m$ and green $a=15\mu m$.

### Section B: Fourier transform of two-dimensional residual dipole field

The dipole field of the ribbon, Fig S2(a), is written as

$$D(x,y) = \begin{cases} Ae^{-ik_0 x \sin\theta} & |x\cos\alpha + y\sin\alpha| < a \\ 0 & |x\cos\alpha + y\sin\alpha| > a \end{cases}, \quad (S1)$$

$A = (\varepsilon_{sample} - \varepsilon_{substrate})E \approx \chi E$ is the same with that in main text, $\theta$ is incident angle of oblique incident angle of exciting laser. If we use new coordinate $(u, v)$

$$\begin{cases} u = x\cos\alpha + y\sin\alpha \\ v = -x\sin\alpha + y\cos\alpha \end{cases}, \quad (S2)$$

then

$$D(x,y) = \begin{cases} Ae^{-ik_0(u\cos\alpha - v\sin\alpha)\sin\theta} & |u| < a \\ 0 & |u| > a \end{cases}. \quad (S3)$$

Similar, the Fourier transform

$$F(k_x, k_y) = \left(\frac{1}{2\pi}\right)^2 \iint D(x,y) e^{-i(k_x x + k_y y)} dx dy, \quad (S4)$$

is written as

$$F(k_u, k_v) = \left(\frac{1}{2\pi}\right)^2 \iint D(u,v) e^{-i(k_u u + k_v v)} du\, dv =$$

$$\left(\frac{1}{2\pi}\right)^2 A \int_{-\infty}^{+\infty} e^{i(k_0 v \sin\alpha \sin\theta - k_v v)} dv \int_{-a}^{a} e^{-i(k_0 u \cos\alpha \sin\theta + k_u u)} du = \left(\frac{1}{2\pi}\right)^2 A \delta(k_v - k_0 \sin\alpha \sin\theta) \frac{2\sin((k_u + k_0 \cos\alpha \sin\theta)a)}{k_u + k_0 \cos\alpha \sin\theta}, \quad (S5)$$

where

$$\begin{cases} k_u = k_x \cos\alpha + k_y \sin\alpha \\ k_v = -k_x \sin\alpha + k_y \cos\alpha \end{cases} \quad (S6)$$

Then

$$F(k_x, k_y) = \left(\frac{1}{2\pi}\right)^2 A \delta(-k_x \sin\alpha + k_y \cos\alpha - k_0 \sin\alpha \sin\theta) \frac{2\sin((k_x \cos\alpha + k_y \sin\alpha + k_0 \cos\alpha \sin\theta)a)}{k_x \cos\alpha + k_y \sin\alpha + k_0 \cos\alpha \sin\theta}. \quad (S7)$$

It is a line, $-k_x \sin\alpha + k_y \cos\alpha - k_0 \sin\alpha \sin\theta = 0$, that perpendicular to the edge in real space, Fig. S2(b).

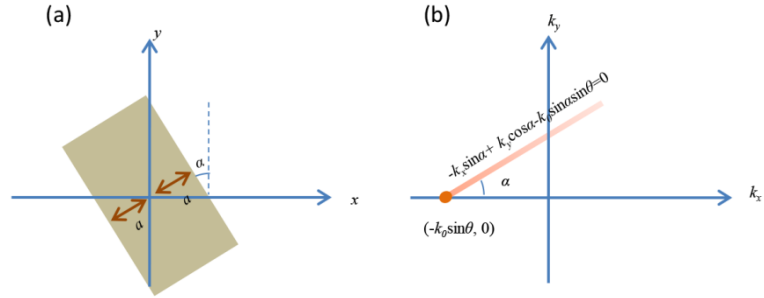

FIG. S2 (a) Coordinate system of tilt ribbon in real space. (b) Sketch of amplitude distribution in k space of $F(k_x, k_y)$.

### Section C: Optical path

Figure S3 shows a detailed schematics of experimental setup. When lamp is on, CCD1 record the reflection image of the sample. For scattering signal measurement, lamp is shut down and laser beam is oblique incident on the sample in xz plane. The reflecting signal is out of optical path, only scattering signal can be collected by the object lens. CCD1 record the scattering image with $100\mu m \times 80\mu m$ field of view. At the same time only the region of interest, with $20\mu m \times 20\mu m$ area, is allowed to pass the aperture that is placed on the conjugate plane of CCD1. Its real and k space images are recorded by CCD2 when the center wavelength of spectrometer is set as 0nm. And also scattering spectra is recorded on CCD2 by rotating the grating.

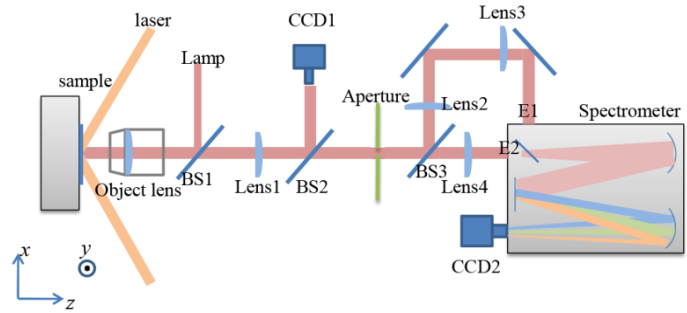

FIG. S3. Schematic drawing of experiment set up. Lamp is reflected by BS1 and passes object lens to illuminate the sample. Laser beam is obliquely incident on the sample out of object lens. Reflection or scattering image is enlarged by object lens and Lens1, and then projection CCD1 and aperture plane at the same time with the help of BS2. Aperture plane is imaged on entrance1 (E1) of spectrometer by Lens2 and Lens3 and is Fourier transformed to entrance2 (E2) of spectrometer by Lens4. Switchable reflector mirror in the spectrometer determines which one enter into CCD2, real space image or k space image.

**Section D: Light scattering light in real space and k space of other samples.**

Figure S4 shows high angle scattering image with different rotation angle $\alpha$. Scattering lines in real space and k space are perpendicular to each other in every sample. And the distances between the zero point of scattering lines all equal to $\sin(66°)\sin(\alpha)$, that according to the theoretical result. In especial, when the rotation angle of two edges of a single flake are all equal to $30^0$, that is close to critical angle $\alpha_c=33°$, then nearly no edge scattering single is collected, in Fig S4 (d-3) and (d-4).

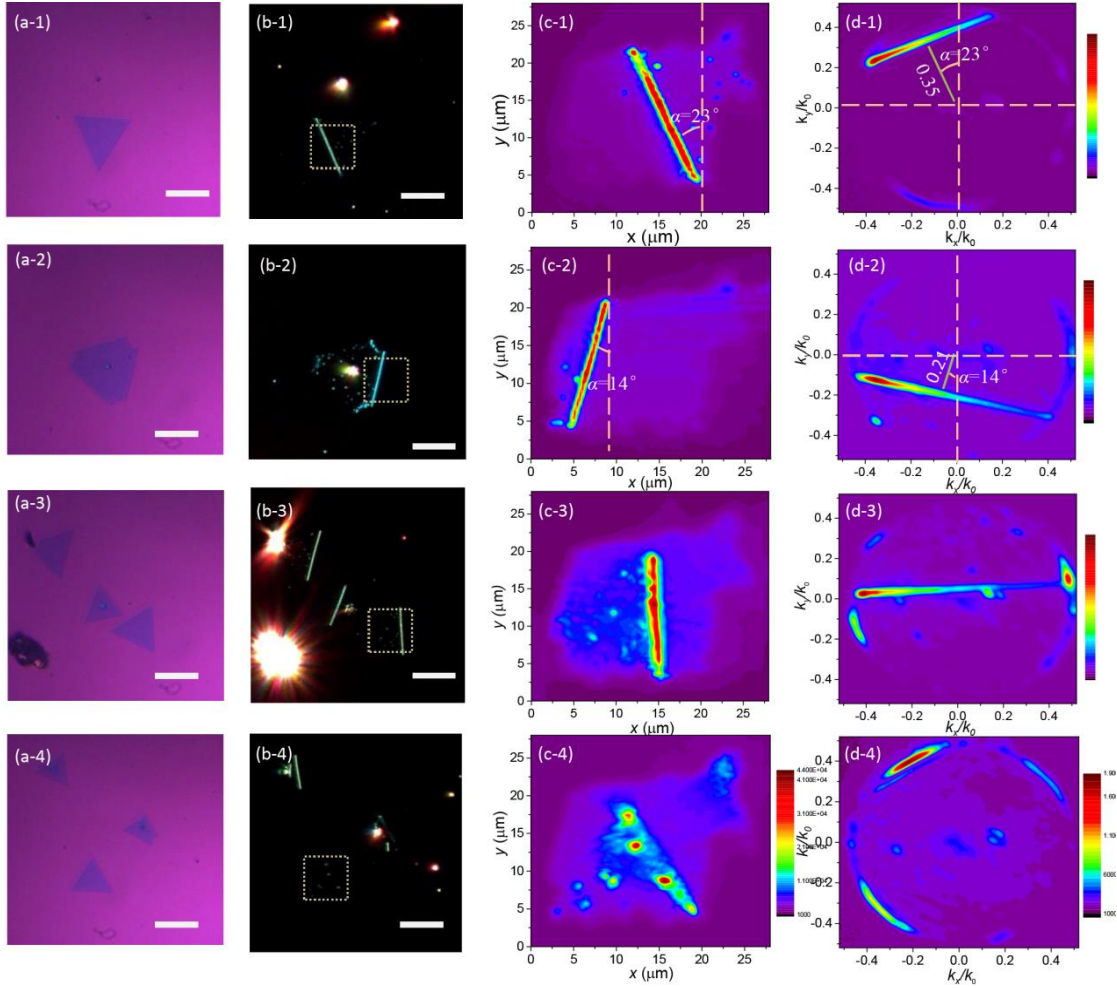

FIG. S4. Reflection and scattering light image of samples with different rotation angle. (a-*i*) the reflection image of *i*th sample. (b-*i*) the scattering light image of *i*th sample in real space recorded by CCD1. All the scale barsare 20 $\mu$m.(c-*i*) the scattering light image that passes through aperture of *i*th sample in real space recorded by CCD2. Aperture allowed area is labeled in (b-*i*) by dot box. (d-*i*) is the scattering light image of *i*th sample in k space recorded by CCD2.

**Section E: Corrected spectra of scattered light with exciting laser spectrum, instrument optical loss, object lens collecting range and dipole polarization.**

The spectra, detected by CCD2 should be corrected by the laser beam spectrum and optical loss spectra of optical element in optical path that are shown in Fig. S5 (a) and (b) respectively. To extract $|\chi_{in}(\omega)|^2$ spectrum, angle distribution correction factor should be calculated first. S wave scattering intensity collected by the object lens is

$$I_s(\omega) = \iint_{\sqrt{k_x^2+k_y^2}<0.5k_0} \gamma \omega^2 \frac{(k_0^2-k_y^2)}{k_0^2} |\chi_{in}(\omega)E_s F'(k_x, k_y)|^2 dk_x dk_y. \quad (S8)$$

Then

$$|\chi_{in}(\omega)|^2 = I_s(\omega)/(\gamma\omega^2 E_s^2 \eta), \qquad (S9)$$

where

$$\eta = \iint_{\sqrt{k_x^2+k_y^2}<0.5k_0} \left(1 - \frac{k_y^2}{k_0^2}\right) |F'(k_x, k_y)|^2 \, dk_x dk_y, \qquad (S10)$$

$\eta$ is angle distribution correction factor that contains the vector correction factor $\left(1 - \frac{k_y^2}{k_0^2}\right)$, and Fourier transform factor $|F'(k_x, k_y)|^2$. Figure S5 (c) shows $\eta$ spectrum for $\theta = 66°$ and $\alpha = 17°$.

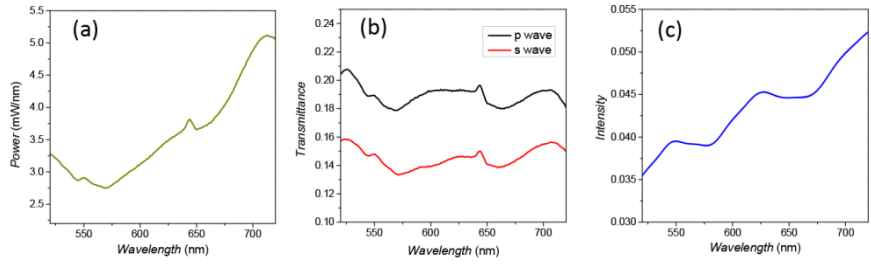

FIG. S5. (a) White laser spectrum. (b) Optical loss spectra of optical element in optical path. (c) Angle distribution correction factor spectrum for $\theta = 66°$ and $\alpha = 17°$.